\documentstyle[aps,prl,twocolumn]{revtex}
\input psfig.tex
\begin{document}
\title{
Oscillating Color Transparency in $\pi A\rightarrow \pi p (A-1)$
and  $\gamma A\rightarrow \pi N (A-1)$}
\author{Pankaj Jain$^a$, Bijoy Kundu$^b$ and John P. Ralston$^c$}
\vskip 0.5cm
\address{\it $^a$Department of Physics, IIT Kanpur, Kanpur-208 016, INDIA\\
$^b$ Institute of Physics, Bhubaneshwar-751 005, INDIA\\
$^c$Department of Physics and Astronomy, University of Kansas,
Lawrence, KS 66045 USA\\
}
\maketitle
\begin{abstract}
{\bf Abstract:} The energy dependence of $90^o$ $cm$ fixed angle
scattering of $\pi p \rightarrow \pi' p'$ and $\gamma p\rightarrow
\pi^+ n$ at large momentum transfer are found to be well described in
terms of interfering short and long distance amplitudes with dynamical
phases induced by Sudakov effects.  We calculate the color
transparency ratio for the corresponding processes in nuclear
environments $\pi A\rightarrow \pi'p(A-1)$ and $\gamma A\rightarrow
\pi N (A-1)$ taking nuclear filtering into account.  A prediction that
the transparency ratio for these reactions will oscillate with energy
provides an important test of the Sudakov phase shift and nuclear
filtering hypothesis which is testable in upcoming experiments.
\end{abstract}

\bigskip
The strong interactions remain a mystery and a phenomenology.  Color
transparency separates conventional strong interaction physics from
perturbative $QCD$.  The perturbative calculation predicts {\it
suppression } of strong interactions in certain exclusive reactions
containing a large momentum transfer $Q^{2}>>GeV^{2}$ subprocess \cite
{brodMuell,PBJ}. Suppression is supposed to occur in initial or final
state interactions with nuclear targets.  The perturbative $QCD$ ($pQCD$)
prediction is dramatic because it apparently contradicts the older
theory in a domain of its validity \cite {brodMuell,PBJ}.  Indeed it 
is not clear whether color
transparency is capable of being described using hadronic coordinates
\cite {jennings93}. At the same time, the many
shortcomings of the $pQCD$ description at the moderate $Q^{2}$
values of experiments \cite{Car,ebeams} 
are well known: Hence the phenomena of color
transparency play a pivotal role from either point of view.

The BNL E850 experiment of Carroll {\it et al}\cite{Car} compared
proton-proton elastic collisions with corresponding quasi-elastic
nuclear processes $pA\rightarrow p'p''(A-1)$.  A transparency ratio
oscillating with energy was observed.  The origin of the oscillations
\cite{RP88,BT88}
remains controversial, and the underlying mechanism has generally been
assumed to be unique to proton-proton reactions.  Here we show that {
\it oscillatory color transparency} is also expected in several
processes involving the pion.  Only recently have the full
perturbative kernels needed for a $pQCD$ description of color
transparency been completely evaluated so that the calculations have a
workable paradigm \cite {etheory,BL}. Here we report calculations predicting
new phenomena observable in experiments currently underway at CEBAF \cite
{gao104}, which will provide fundamental information on how pQCD may
be applied to exclusive processes both in free space and in a nuclear
medium.  Other experimental predictions can be checked at BNL or other
hadron beam laboratories.  The predictions are rather distinctive, and
tests of the entire framework of color transparency become available.

Consider the reaction $\pi p\rightarrow m' p'$ compared to $\pi
A\rightarrow m' N' (A-1)$, where $m'$ represents a meson and $N'$
represents a nucleon.  These processes contain Landshoff pinch
singularities, and in $pQCD$ are expected \cite{Landshoff} to show
oscillations about power-law energy dependence at fixed angle.  The
wavelength of the oscillations (imaginary anomalous dimensions) are
calculable, but have not yet been calculated.  These important calculations
have only been performed so far for selected diagrams occuring in $pp$
scattering \cite{StermanBotts}.  Now considering the $\pi p$ reactions, the
same physics of pinch singularities again demands a factorization scheme
more general than the short-distance ``quark-counting" method\cite{GPL},
which is sometimes misunderstood to define $pQCD$.  We re-iterate that among
the various competing renditions --the short distance factorization of
Brodsky-LePage, the asymptotic impact-parameter factorization of Botts and
Sterman, and the finite $Q^{2}$ factorization of Gousset {\it et al}
\cite{gousset} incorpating spin-effects, all represent $pQCD$ and
concepts
such as ``asymptotic'' or ``short distance" are not synonymous with ``$pQCD$''.

To represent all the diagrams and integration regions, our
calculations include the transverse spatial separation $b$ between
quarks \cite {StermanBotts,RP90}.  Several remarkable things emerge:
First, the naive association of $b\sim 1/Q$ breaks down, and hard
reactions depend on the entire region $ 1/Q< b< 1/\Lambda_{QCD} $.
There is generic violation of the short-distance selection rule known
as ``hadron-helicity conservation'', a model-independent test of the
short-distance framework \cite{helicity81}, and $pQCD$ predicts
non-trivial transverse and helicity-violating spin effects for large
$Q^{2}>>GeV^{2} $ \cite{Landshoff}.  Next, Sudakov factors regulate
the approach to the pinch configurations and must be included among
the kernels.  The Sudakov factors reinstate the geometrical
strong-interaction $Fm$-scale by drastically cutting off amplitudes at
distances larger than $1/\Lambda_{QCD}$.  The Sudakov-improved amplidues
must obey analyticity, exhibited in $pQCD$ by color and
flavor matrix phase factors of the form $\exp[-i \pi c \ln(\ln(s/
\Lambda^{2}_{QCD}))]$, handled by extending the notion of anomalous
dimensions to purely imaginary numbers\cite{pire82}.  In a nuclear medium
large-$b$ regions interact inelastically  with
exponential attenuation, while those regions of small $b$ interact
proportional to $b^{2}\rightarrow 0$, resulting in transparency
\cite{low}. 
By depleting the long distance amplitudes, ``nuclear filtering''
quantum mechanically favors short distance processes in large nuclei 
\cite {RP90,low}.

All of these elements are seen in free-space $pp$ reactions and the
$BNL$ color transparency experiment.  In particular the free space
cross section $s^{10} d\sigma/dt$ at fixed $cm$ angle $90^{0}$
oscillates with $\ln(\ln(s/ \Lambda^{2}_{QCD}))$.  (Here $s,t$ are
the Mandlestam variables for $cm$ energy-squared and momentum
transfer-squared).  The color transparency ratio was found to show
oscillations $180^o$ out of phase with the free-space oscillations.
In a two-component model this rather unambiquously indicates strong
attenuation or ``filtering'' in the nuclear medium of one
long-distance amplitude, and little attenuation of another
short-distance component.  In addition, attenuation cross sections
extracted from the data \cite {jain93} are substantially smaller than
the traditional 40 $mb$ of conventional strong interaction physics at
these energies, and scaling in the variable $Q^2/A^{1/3}$ was
observed.  Consistently, the cross section in the nuclear target shows
negligible oscillations with energy \cite{Car} and apparently conforms to
predictions of short-distance physics \cite{GRF,jain93}.  In
contrast, a model based on the hadronic basis (Farrar { \it et al }
Ref. \cite{jennings93}) fails to describe the data by many
standard deviations.

Correlating these observations of $pp$ reactions with the dynamical
similarity of $\pi p$ reactions suggests similar phenomena should be
observed.  Parts of the calculations are stymied by a major difficulty: no
systematic method exists to find the relative phases of exclusive
amplitudes.  It is not enough to calculate the phase of the
asymptotically largest amplitude (the procedure of Ref.
\cite{StermanBotts}) but it is necessary to find any sizable phase
coefficient of any sizable amplitude.

As a practical resolution we have fit the $90^{o} cm$ fixed angle
$s^{8 }d\sigma/dt$ data \cite{owen} for $\pi p\rightarrow \pi' p'$
with a two-component model.  The existence of this data for fixed
angle scattering compiled by Blazey \cite{owen} (Fig.  1a) appears not
to be widely appreciated.  Oscillations in this data show much the
same features as the free-space $pp$ data.  With $M$ denoting the
$2\rightarrow2$ amplitude for the reaction, our fit is given by:
\begin{eqnarray}
s^8{d\sigma\over d|t|} = \bigg|A_0 &+& {A_1\sqrt s e^{-ic_1\log\log 
q^2/\Lambda_{QCD}^2}
\over (\log s)^{d_1}}\nonumber\\
&+& {A_2e^{-ic_2\log\log q^2/\Lambda_{QCD}^2}\over (\log s)^{d_2} }
\bigg|^2
\end{eqnarray}
  where $A_0,A_1,A_2,c_1,c_2,d_1,d_2$ are real parameters.  The
  functional form of our exponents have been updated compared to Ref. 
\cite{pire82} and come from expanding the imaginary parts of
  Sudakov exponents.  In accord with the discussion, the two
  components, $A_1$ and $A_2$ represent regions of large $b$,
  associated Sudakov effects, and logarithmically-varying phases, while
  small-$b\sim 1/Q$ regions are described by short-distance theory.
  The best fit gives $A_0 = -0.638, A_1=5.1, c_1 = 25.6, d_1 = 5.13,
  A_2 = -0.065, c_2 = -26.3, d_2=-1.16$ with $\chi^2/dof = 1.97$.  If we
  include only one long distance amplitude setting $A_2 = 0$, then the
  best fit gives $A_0 = -0.661, A_1=-7.67, c_1 = 23.2, d_1 = 6.03$ with
  $\chi^2/dof = 5.01$.  
In
  comparison the short-distance model $s^{-8}$ fit gives $\chi^2/dof =
  99$.

Now turn to the corresponding pion-initiated reaction with a nuclear
target.  In the two-amplitude model each component interacts with the nuclear
target by a different rule. For the long distance pieces, the target measures
the integration region (``transverse size") via attenuation by the 
rule $I_{j} = exp(- \int k \sigma_{j}
n dz)$ where $z$ is the straight-line propagation distance across the
target, and $n$ is the nucleon density; $\sigma_{ab}$ is the
absorptive cross section for particles $a,b$. We used $\sigma_{pp}=40
mb$, and $\sigma_{\pi p}=26mb$.  The short
distance amplitude is attenuated with a model inspired by 
$\sigma_S=k/(x_1x_2 Q^2)$, where
$x_1,x_2$ are the momentum fractions of the quarks inside the proton.
Since this amplitude is short-distance we set $x_1=x_2=0.5$ and so
$\sigma_S= k(1.6 mb)  (GeV^2/Q^2) $. Short-range nuclear
correlations are included \cite{MS} in both cases.  We then calculate 
the cross section in the nuclear case by
\begin{eqnarray}
s^8{d\sigma_A\over d|t|} &=& {1\over A} \int d^3x n(x)\bigg|A_0I^{i}_{S}
I^{fa}_{S}I^{fb}_{S}\nonumber\\ &+&
{A_1\sqrt s e^{-ic_1\log\log q^2/\Lambda_{QCD}^2+i\phi_A}
\over (\log s)^{d_1}}I^i_{\pi p}I^f_{\pi p}I^f_{pp}\nonumber\\
&+& {A_2e^{-ic_2\log\log q^2/\Lambda_{QCD}^2+i\phi_A}
\over (\log s)^{d_2} }I^i_{\pi p}I^f_{\pi p}I^f_{pp}
\bigg|^2
\end{eqnarray}
where $A$ is the nuclear number; superscripts $i$ and $f$
refer to initial and final state attenuation factors, respectively. 
The formula indicates we
took into account a potential relative phase $\phi_{A}$ between the
two amplitudes due to interaction with the nucleus.

We assume Fermi-motion is taken out experimentally by overdetermined
kinematic reconstruction (such as possible at $BNL$) and so this has
not been included in the calculations.  We treat $\phi_{A}$ and $k$ as
parameters subject to considerable uncertainty.  However for the
entire range of $0< \phi_{A}<2\pi$ and varying $5<k< 10$ the
calculations are sufficiently robust to predict rather dramatic
effects.  In Fig. 1b we show the results for the transparency ratio,
$T = d\sigma(\pi A\rightarrow m'N' (A-1); 90^{o})/dt/
\left[Zd\sigma(\pi p\rightarrow \pi p'; 90^{o}) /dt\right] $ for the
two different models.  The plots (Fig 1, b-c) show a striking
$180^{o}$ phase shift between the oscillations of the transparency
ratio and those seen in the free-space reaction.  $T$ is less
sensitive to variations of $\phi_{A}$ compared to $k$: for all values
of the $\phi_{A}$ we find that $T$ shows significant oscillations with
energy.  Only for very large values of $k>> 10$ do these oscillations
disappear, a limit in which no short distance contribution effectively
exists.  The plots are given for large nuclei where the calculation
indicates filtering will be effective: for $A>>1$, short distance
physics predicts scaling in the variable $Q^2/A^ { 1/3 }$.
The theory may be extended to
smaller $A\approx 12$, where our calculations also show a substantial
effect, with less confidence regarding the importance of the
short-distance component.  As in the $pp$ case, the measurement of the
transparency ratio as a function of $s$, or the $A$ dependence at
fixed large $s$, would be capable of ruling out the hadronic-basis
predictions for the same reaction, which are either monotonic (Glauber
theory) or linear in the energy (exploding point-like classical
expansion theory (Farrar {\it et al }, \cite{jennings93}).

\begin{figure}
\psfig{file=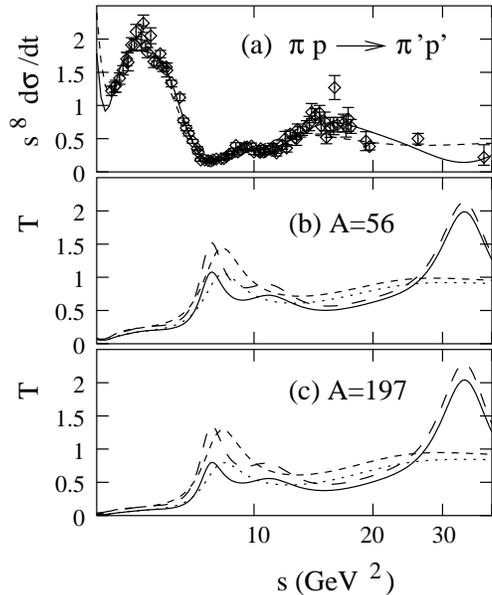}
\caption{(a) The free space $\pi p$ $90^o$ cross section
$s^8 d\sigma/d|t|$ ($10^8$ GeV$^{16}$ $\mu$b/GeV$^2$) using three component
model (solid curve) and a two component model (dashed).  (b,c)
Calculated color transparency ratio for $A=56,197$ using nuclear
filtering, in the three component model with $k=10$ (solid), $k=5$
(long dashed) and the two component model with $k=10$ (dotted) and
$k=5$ (short dashed).} 
\end{figure}

Most exciting are experimental data and rapidily upcoming prospects
for the processes $\gamma p \rightarrow \pi^+n$ and $\gamma
n\rightarrow \pi^- p$.  Data exists for $s < 16\ GeV^2$ and $s<4\
GeV^2$, respectively \cite{anderson}.  The Jefferson Lab and CEBAF is
soon expected to extend the energy range of $\gamma n$ reactions to
about $s=16$ GeV$^2$ with high precision, as well as measure the color
transparency ratio for this process \cite{gao104}.  The short distance
theory predicts $d\sigma/dt_{90^{o}}\sim s^{-7}$, within which
framework it has been shown for asymptotically large momentum transfer
\cite{FSZ} that Landshoff pinches are absent.  Unlike $pp$ and $\pi p
$ reactions, then, where the pinch regions actually constitute the
asymptotic prediction, here the Landshoff and associated Sudakov phase
physics is subleading.  But the asymptotic limit (infinte energy) has
little weight for laboratory $Q^{2}$, and ubiquitous
observations of spin-effects forbidden by
short-distance theory but part of regular $pQCD$ \cite {gousset} argue
for the more complete treatment including the pinches.  Most
interestingly, the existing data show considerable oscillations around
power dependence (Fig.  2a).  Like the $\pi-p$ case, the existence of
this data also appears not to be widely appreciated.

We fit the experimental data for $\gamma p\rightarrow \pi^+n$ with
center of mass scattering angle $90^o$ and $\sqrt s>2 $ GeV. The best
fit to the 17 data points available is shown in Fig. (2).  We use the
same amplitude ansatz as Eq (1) for $s^7 d \sigma/dt$, obtaining $A_0 =
0.90, A_1 = 2.65/s,
c_1 = 64.5, A_2 = 8.01/s, c_2 = -126.4$ with $\chi^2/dof 
= 0.69$.  Here
we have set $d_1=d_2=4$, as the quality of fit does not depend
substantially on these parameters.  The values of $d_1$ and $d_2$ were
chosen to obtain a relatively flat free space behavior beyond $\sqrt
s=3.0$ GeV, where the presence or absence of oscillations remains
experimentaly unstudied.  We arbitrarily imposed a model of
short-distance physics for this region.  If we set $A_2=0$ then the
best fit gives, $A_0=0.89$, $A_1=-4.15$ and $c_1=79.8$ with $\chi^2$
per degree of freedom of 1.09.  For comparison the short-distance
$s^{-7}$ model gives $\chi^2/dof =2.9$.  While our fit is favored
statistically, including effects of extra parameters, the
short-distance model is not ruled out in comparison.  Cutting the
experimental uncertainties in half would be pivotal.  We mention this
because the uncertainties are expected to decrease with the
experiments imminent.

For the nuclear process $\gamma A\rightarrow \pi^+n+(A-1) $, we calculate
$s^7{d\sigma\over d|t|}$ with the same format as Eq (2).  Results for the
transparency ratio for $A=12,56,197$ are shown in Fig.  2.  In calculating
filtering factors we conservatively assume that the incident photon does
not attenuate significantly.  While there are many models to attenuate the
photon somewhat, this allows a conservative presentation, because the
effects of filtering which generate the oscillating transparency ratio are
minimized. Let us note that experimentally the final state
$N$ can be a proton or a neutron, but to predict the neutron case
definitively we would
need free space neutron scattering data that we do not currently have.

Observing Fig.  (2), the predicted transparency ratio $T$ again oscillates
180$^o$ out of phase with the free space cross-section.  This simple fact
has so far not been appreciated as generic, and previous hadronic-basis
estimates for the transparency ratio have not taken the oscillations in
free space data into account, yielding monotonically increasing energy
dependence\cite{gao104}.  The upcoming photon-initiated experiments, 
then, may be on
the verge of confirming a third case of oscillating fixed angle data, and
oscillating color transparency.

Color transparency with a photon beam remains significantly different
from hadron initiated processes.  The distinction becomes clear when
the $Q^2$ dependence of a virtual photon is used as an experimental
tool.  In the limit of large $Q^2>>GeV^2$, experimental evidence from
deeply-inelastic scattering overwhelming supports the concept of a
point-like photon interaction, with negligible attenuation and
pertubatively understood hadronic components in scattering.  The lack
of pinch singularities of the large-$Q^2$ framework predicts power-law
fading of oscillations in both free space cross section and
transparency ratio in the limit of large-$Q^2$. 
The regime of
large-$Q^2$ for photons should coincide with the regime of Bjorken
scaling, so that the moderate $Q^2$ of existing electron beams should
suffice.  This would be extremely interesting and productive area to
explore experimentally.  

\begin{figure}
\psfig{file=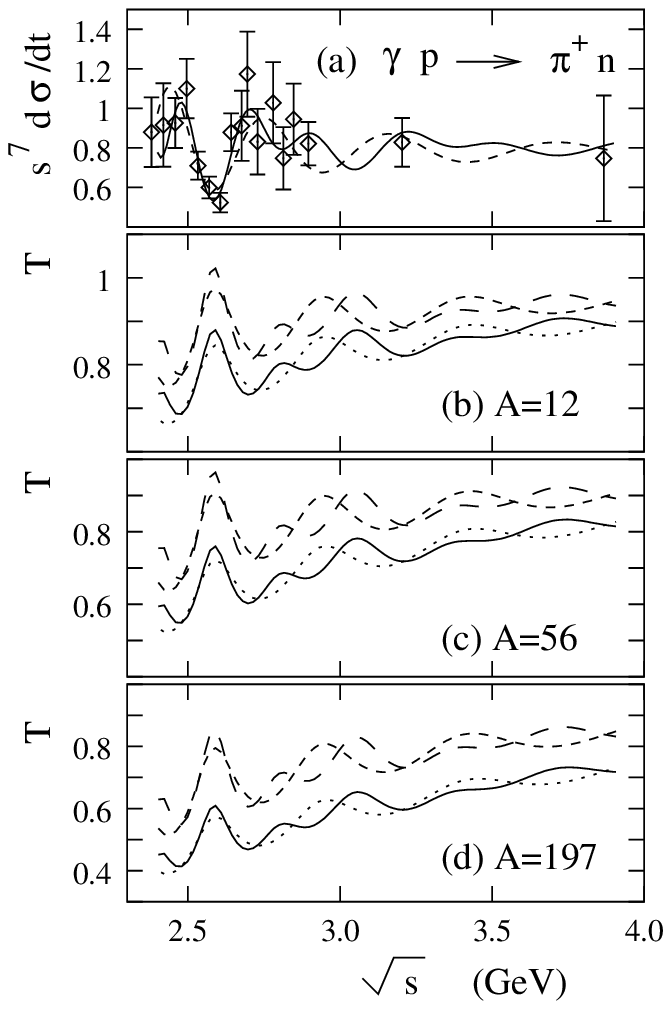}
\caption{(a) The free space $\gamma p
\rightarrow \pi^+ n$ $90^o$ cross section $s^7d\sigma/d|t|$ ($10^7$
GeV$^{14}$ nb/GeV$^2$) using three component model (solid curve) and a
two component model (dashed).  (b,c,d) Calculated color transparency
ratio for $A=12,56,197$ using nuclear filtering, in three component
model with $k=10$ (solid), $k=5$ (long dashed) , and in two component
model with $k=10$ (dotted) and $k=5$ (short dashed).
}
\end{figure}

Direct tests of the hadron-helicity non-conserving character of the
pinch-singularity regions are very interesting.  A pion beam suggests
studying reactions involving a final-state $\rho$ meson: again the
process has pinch singularities.  The $pQCD$ analysis indicates that
oscillations of fixed-angle scattering with energy will occur, and
indeed one of the points of this paper is that such oscillations are
{\it generic}.  The failure of short-distaince models, and dynamical
importance of the pinch regions for $\pi p \rightarrow \rho p$ is
supported by observations \cite {Heppelrhos} of final-state
$\rho$-polarization density matrix elements $\rho_{1,-1}$ of order
unity.  If this is due to the pinch regions, as expected \cite
{Landshoff}, then filtering
in a large nucleus should remove them.  Oscillating polarization
effects would be very dramatic: $\rho_{1,-1}$ oscillating with energy
at fixed angle is expected if the dynamical phases are correlated with
exchange of orbital angular momentum.  Counting powers of the internal
coordinate $b$ and the units of orbital angular momentum, we can
predict that at fixed large $ Q^{2} $, each power of $b^{2}$ in
amplitude calculations will scale like $A^{-1/3}$ due to nuclear
filtering.

To conclude, oscillating color transparency is a generic prediction of
$pQCD$, testable with imminent experiments.  
We believe that the
observation of oscillations in experimental data for the transparency
ratio, consistently $ 180^o $ out of phase with the free space
counterparts, and in three independent reactions, will be strong
confirmation of nuclear filtering and the basic $pQCD$ understanding
of color transparency.

\bigskip \noindent {\bf Acknowledgements} We thank Haiyan Gao for very
useful discussions and for providing data for $\gamma p$.  We also
thank Jerry Blazey for help with data and Bernard Pire for helpful
comments.  This work was supported in part under the Department of
Energy, and the {\it Kansas Institute for Theoretical and
Computational Science/ K*STAR} program.

\end{document}